\documentclass[amsmath,showpacs,showkeys,superscriptaddress]{revtex4}

\usepackage{graphicx,multirow,array}
\usepackage{bm,hhline}

\newcommand{\be}{\begin{equation}}
\newcommand{\ee}{\end{equation}}
\newcommand{\bea}{\begin{eqnarray}}
\newcommand{\eea}{\end{eqnarray}}
\newcommand{\hf}{\frac{1}{2}}

\newcommand{\he}{$^4$He}

\def\eq#1{(\ref{#1})}
\def\mathrm{}

\def\cK{{\cal K}}
\def\cM{{\cal M}^2}
\def\P{\hat P^2_+}
\def\p2{\hat P^2}

\def\mr#1{{\mathrm{#1}}}
\def\ord#1{{\cal O}\left(#1\right)}

\begin{document}
\title{Effective field theory for $^4$He}
\author{J.M.~Carmona}
\email{jcarmona@unizar.es}
\affiliation{Departamento de F\'{\i}sica Te\'orica, Universidad de
Zaragoza, Zaragoza, Spain}
\affiliation{Instituto de Biocomputaci\'on y  F\'{\i}sica de
Sistemas Complejos (BIFI), Corona de Arag\'on 42, Zaragoza 50009, Spain}
\author{S.~Jim\'enez}
\email{sergiojs@unizar.es}
\affiliation{Dipartimento di Fisica, INFM and SMC, U. di Roma {\em La Sapienza},
P.le A. Moro 2, Roma I-00185, Italy.}
\affiliation{Departamento de F\'{\i}sica Te\'orica, Universidad de
Zaragoza, Zaragoza, Spain}
\affiliation{Instituto de Biocomputaci\'on y  F\'{\i}sica de
Sistemas Complejos (BIFI), Corona de Arag\'on 42, Zaragoza 50009, Spain}
\author{J.~Polonyi}
\email{polonyi@fresnel.u-strasbg.fr}
\affiliation{Laboratory for Theoretical Physics, Louis Pasteur University,
Strasbourg, France}
\affiliation{Department of Atomic Physics, Lorand E\"otv\"os University,
Budapest, Hungary}
\author{A.~Taranc\'on}
\email{tarancon@sol.unizar.es}
\affiliation{Departamento de F\'{\i}sica Te\'orica, Universidad de
Zaragoza, Zaragoza, Spain}
\affiliation{Instituto de Biocomputaci\'on y  F\'{\i}sica de
Sistemas Complejos (BIFI), Corona de Arag\'on 42, Zaragoza 50009, Spain}
\date{\today}
\begin{abstract}
We introduce an effective scalar field theory to describe
the \he\ phase diagram, which can be considered as a generalization of the
XY model which gives the usual $\lambda$-transition. This theory results from
a Ginzburg-Landau Hamiltonian with higher order derivatives, which allow to
produce transitions between the superfluid, normal liquid and solid phases
of \he. Mean field and Monte Carlo analyses suggest that this model is able to
reproduce the main qualitative features of \he\  phase transitions.
\end{abstract}


\maketitle

\section{Introduction}
\label{intro}

Effective field theory models are widely used to describe phase
transitions in condensed matter systems. The simplest example is
that of the Ginzburg-Landau theory~\cite{GL}, in which a
``Hamiltonian'' $H_{\text{GL}}[\phi_i]$ depending on certain
field variables $\phi_i$ and defined on the sites $i$ of a
lattice or at points $x$, $\phi(x)$, in its continuum formulation,
may account for the description of first or second-order phase
transitions.

The effective theory is not intended to be a microscopic theory at
all. Instead, it makes use of the fact that critical phenomena are
divided into universality classes which are determined by a
few basic properties of the system only, such as the dimensionality
of the space, the range of interactions, the number of
components and the symmetry of the order parameter. The
renormalization-group theory predicts that, within a given
universality class, the critical exponents and the scaling
functions are the same for all systems so that we can make use of
the corresponding simpler effective theory to calculate such
quantities.

In this context, the superfluid transition of \he, occurring along
the $\lambda$-line $T_\lambda(P)$, where $P$ is the pressure and
$T$ the temperature, belongs to the three-dimensional XY
universality class~\cite{xy}.
Its order parameter is related to the complex quantum amplitude
of helium atoms~\cite{GLhelium}, so that the O(2) field theory
may serve as an effective description of this transition.
In fact very good agreements between the critical exponents and scaling
functions of this model and the experimental measurements of the
$\lambda$-transition of \he\ are found~\cite{pelissetto}.

\begin{figure}[tb]
\begin{center}
\includegraphics[width=0.9\columnwidth]{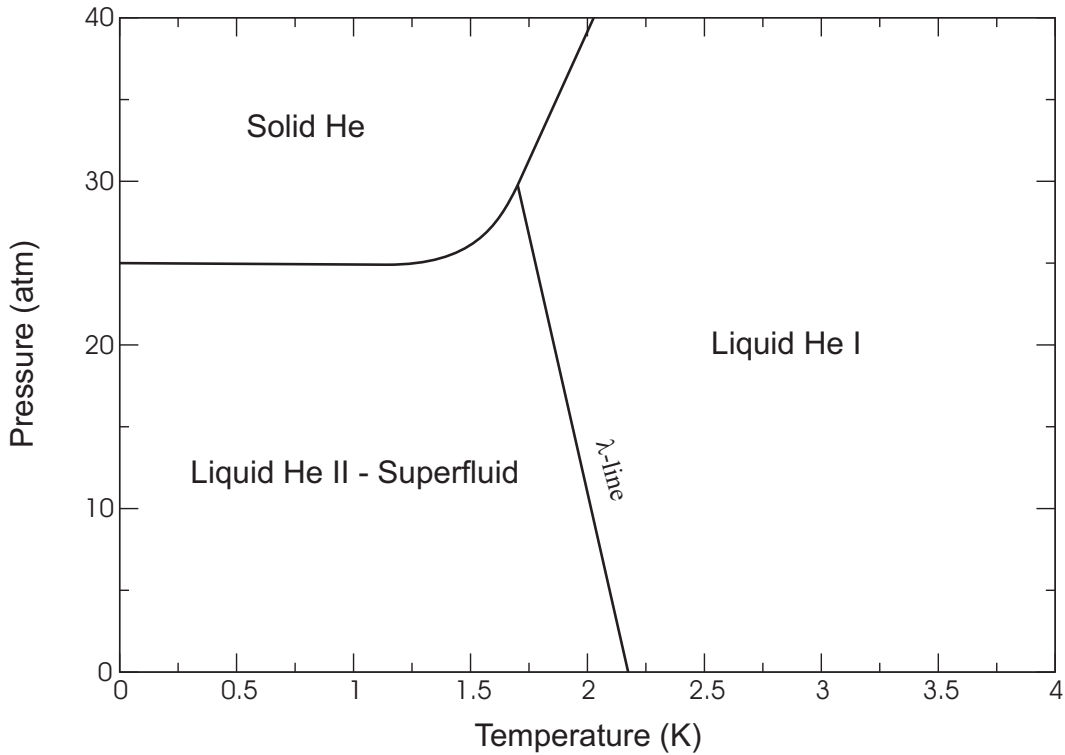}
\caption{Experimental pressure-temperature phase diagram for \he \ in
the low temperature region. Two liquid phases, He I and superfluid He
II, and a solid phase can be distinguished.} \label{fig:helio}
\end{center}
\end{figure}

Fig.~\ref{fig:helio} shows the \he\ phase diagram in the $(T,P)$
plane. The O(2) field theory is intended to describe this system
in the vicinity of the $\lambda$-transition, and strictly speaking
only near $T_c$ at zero pressure, because the model presents a
temperature-driven transition only. The whole $\lambda$-line is
however expected to belong to the same universality class for
$P\neq 0$.
The transition lines between the liquid phases and the
solid phase are experimentally observed to be of first order, with a
finite entropy difference between the phases and the presence
of a latent heat~\cite{wilks}.

In Fig.~\ref{fig:helio}, the solid, superfluid and normal liquid
phases meet at a single point, suggesting the presence of a
Lifshitz point~\cite{LP} in the \he\ phase diagram. In order to
examine the critical behavior around this point, it would be very
convenient to have an effective theory of this system containing
all the three phases which meet there. This model should then be
able to account for the transitions between the solid and the
superfluid phases, solid and normal liquid phases, and superfluid
and normal liquid phases.

We address the problem of finding such an effective theory in the
present work. We take as an starting point the O(2) field theory,
which describes the transition between the superfluid and normal
liquid phases. In this theory, the superfluid phase corresponds to a
\emph{ferromagnetic} phase, in which there exists long-range order,
while such an order is absent in the normal liquid phase, which
corresponds to an unordered or \emph{paramagnetic} phase. In order
to generalize the O(2) field theory to a model including also a
solid phase we should extend the regime of applicability of the
effective theory towards shorter length scales below the lattice
spacing of the solid.

What is the salient feature of the interactions at distances
comparable to this lattice spacing? The periodic ordering in the
ground state underlines the importance of the distance dependence of
the forces acting in this scale regime. What kind of terms of the
effective theory can reproduce an important distance dependence? The
ultralocal polynomials of the field, without space-time derivatives,
represent momentum (i.e.~distance) independent interaction strength.
When such a vertex is inserted into a graph then the distance
dependence of the propagators can induce distance dependent vertex
functions and therefore forces. But this is an indirect mechanism,
the distance dependence actually originates from the propagators,
the usual $\ord{p^2}$ part of the action. When the vertices carry
explicit momentum dependence then the resulting forces should
display more pronounced distance dependence. Therefore one suspects
that terms with higher order derivatives might be the key to
reproduce the solid phase.

The simplest vertex with higher order derivatives is quadratic in
the field variable, meaning the modification of the free dispersion
relation. If the dispersion relation turns out to be negative at a
certain momentum, then the elementary excitation with such a
momentum starts to condense in the vacuum. This condensation will be
stopped and the system will be stabilized by the repulsive
$\ord{\phi^4}$ interaction. Starting from the normal phase, the
appropriately chosen mass term or the coefficient of the higher
order derivative term will take us into the superfluid or the solid
phase which will be called \emph{modulated} phase. Note that the
effective model may also contain an \emph{antiferromagnetic} phase,
in which the period length is the shortest possible scale of the
theory. Such a phase will be however non-physical, since effective
theories are supposed to give sensible physical descriptions only at
energies lower than their energy cutoffs.

The universality argument of the normal-superfluid transition can be
extended towards the solid phase at least in the leading order of
the perturbation expansion. In fact, the perturbative, one-loop
renormalizability at the Lifshitz-point \cite{bmp} provides us the
universality at this order. It remains to be seen if the
universality can be established nonperturbatively.

The heuristic reasoning above applies to any system with modulated
ground state, such as the usual solid state crystal, Wigner-lattice
of dilute electron gas and charge density wave state. But there is
another consideration which makes the argument plausible in the
context of \he, namely the existence of rotons. The higher order
derivative terms of the effective action are present in either
phases, their strength varies only when the phase boundaries are
crossed. As we will see, the existence of rotons, the enhancement of
elementary excitations at a given momentum which corresponds to a
local minimum in the dispersion relation, is naturally reproduced by
means of an action whose quadratic part in the field contains higher
order derivatives. In fact, the rotons should correspond to the
local minimum of the dispersion relation which becomes the absolute
one as the superfluid-solid transition is crossed.

It is interesting to notice the formal similarity between the
dynamics which drive the normal-to-superfluid and the
normal-to-solid transitions. The former is spontaneous symmetry
breaking where the potential energy reaches its minima at
nontrivial, i.e.~nonsymmetrical values of the field, at a
nontrivial scale in the internal space. The symmetry
spontaneously broken is an internal one and the dynamics is modified
mainly in the IR domain. Similar phenomenon may take place in the
external space, which leads to the modulated
phase. If the dispersion relation reaches its minimum at
nonvanishing momentum then particles with such momentum condense
and the vacuum becomes modulated. This mechanism is driven by
the derivative terms and modifies the dynamics mainly at
length scales comparable with the inverse momentum of the
particles condensing. The symmetry broken dynamically is an
external one, rotations and translations.

The organization of our paper is the following. In Section~II the
effective theory for helium-$4$ is defined both in mathematical
and physical terms. Section~III is
devoted to the mean field solution of the model. In Section~IV
we compare the mean field predictions with numerical Monte Carlo
simulations of the system. We analyze its phase diagram and
investigate the order of the transitions between phases. Finally,
the conclusions are presented in Section~V.

\section{Effective theory}

Our model is an extension of the Ginzburg-Landau model in three
dimensions for a complex scalar order parameter by adding higher order
derivatives to the action,
\be
S_c[\phi]=\int_x\left[\hf\partial_\mu\phi_j(x)\cK(a^2\partial^2)
\partial_\mu\phi_j(x)+\hf m^2_c\phi_j(x)\phi_j(x)
+\frac{\lambda_c}{4!}(\phi_j(x)\phi_j(x))^2\right],
\label{GLaction}
\ee
where $j=1,2$, there is an implicit summation over repeated indices, and
the kinetic energy contains the function
\be
\cK(z)=1+c_2z+c_4z^2.
\ee
The theory is regularized on a lattice of spacing $a$ (the ultraviolet
cutoff, which will be taken to be smaller than the period of the solid phase),
so that the integral becomes $\int_x\to a^3\sum_x$.
One uses the dimensionless lattice field variable $\phi_x=\sqrt{a}\phi(x)$,
and gets
\be
S[\phi]=\sum_x\left[-\hf\phi_{j,x}\Delta\cK(\Delta)
\phi_{j,x}+m^2\phi_{j,x}\phi_{j,x}+\frac{\lambda}{4!}
(\phi_{j,x}\phi_{j,x})^2\right],
\label{lattaction}
\ee
with
$\Delta\phi_x=\sum\limits_{\mu=1}^3(\phi_{x+\hat\mu}+\phi_{x-\hat\mu}-2\phi_x)$,
$m^2=m_c^2a^2$, $\lambda=a\lambda_c$.

Expanding $\Delta\cK(\Delta)$ in $S[\phi]$, we obtain
\be
S[\phi]=-\kappa_1 S_1-\kappa_2 S_2 - \kappa_3 S_3 - \kappa_4 S_4
- \kappa_5 S_5 - \kappa_6 S_6 + \kappa \sum_x \phi_{j,x}\phi_{j,x}
+\frac{\lambda}{4!}\sum_x \left(\phi_{j,x}\phi_{j,x}\right)^2,
\label{laactionred}
\ee
with the following definitions for the different terms and coefficients:
\begin{eqnarray}
S_1=\sum_{x,\mu} \phi_{j,x}\phi_{j,x+\hat\mu}\, , &\quad&
\kappa_1=1-12\, c_2 + 123 \,c_4\, , \label{kappa1} \\
S_2=\sum_{x,\mu} \phi_{j,x}\phi_{j,x+2\hat\mu}\, , &\quad&
\kappa_2=c_2-18 \,c_4\, , \\
S_3=\sum_{x,\mu} \phi_{j,x}\phi_{j,x+3\hat\mu}, &\quad&
\kappa_3=c_4\, , \\
S_4=\sum_{x,\mu<\nu} \phi_{j,x}\left(\phi_{j,x+\hat\mu+\hat\nu}+
\phi_{j,x-\hat\mu+\hat\nu}\right), &\quad&
\kappa_4=2 \,c_2 - 36 \,c_4\, , \\
S_5=\sum_{x,\mu<\nu} \phi_{j,x}\left(\phi_{j,x+2\hat\mu+\hat\nu}+
\phi_{j,x-2\hat\mu+\hat\nu}+\phi_{j,x+2\hat\nu+\hat\mu}+
\phi_{j,x-2\hat\nu+\hat\mu}\right), &\quad&
\kappa_5= 3\, c_4\, , \\
S_6=\sum_x \phi_{j,x}\left(\phi_{j,x+\hat 1+\hat 2+\hat 3}+\phi_{j,x+\hat 1-\hat 2+\hat 3}
+\phi_{j,x-\hat 1+\hat 2+\hat 3}+\phi_{j,x-\hat 1-\hat 2+\hat 3}\right), &\quad&
\kappa_6= 6\, c_4\, , \\
 &\quad& \kappa=\frac{m^2}{2}+3-21\, c_2 + 162\, c_4\, . \label{kappa}
\end{eqnarray}

\subsection*{Physical interpretation}
\label{physint}

This effective model presents a number of parameters: $m^2$, $\lambda$,
$c_2$, and $c_4$. $m^2$ and $\lambda$ offer the possibility to have
a broken symmetry, as in the standard Ginzburg-Landau Hamiltonian. Indeed,
the polynomial
\be
H(\varphi)=\frac{1}{2}r_0\varphi^2+\frac{1}{4!}u_0 \varphi^4, \quad u_0>0
\ee
has two minima for $r_0<0$. The coefficient $u_0$ must be positive
so that $\lim_{\varphi\to\pm\infty} H(\varphi)=+\infty$, which guarantees the
stability of the minimum of $H(\varphi)$.

Therefore, $m^2<0$ and $m^2>0$ will give the broken and the paramagnetic
phases, respectively, in the tree-level approximation. The parameter
$m^2$ has to vanish linearly at the critical temperature,
$m^2\propto (T-T_c)$.

The other two parameters, $c_2$ and $c_4$, appearing in the higher
order derivative part of the action, will be responsible for the
emergence of the solid phase. Since making $c_2,c_4\to 0$ in
Eq.~\eqref{GLaction} gives the usual O(2) field theory, which
describes the \he\ transition at $P=0$, we will associate the
pressure to a linear combination of these two parameters, to be
determined afterwards.

\section{Mean field solution}
\label{mfsol}

The first, simplest step in determining the phase structure of the
model is the tree-level solution which produces an inhomogeneous
mean field in our case. Since this model is translational and
rotational invariant in space we shall look for a mean field
vacuum of the form
\be\label{trconf}
\begin{pmatrix}
\phi_{1,x}\cr\phi_{2,x}
\end{pmatrix}
=\phi\begin{pmatrix}
\cos\left(K^\mu x^\mu\right)\cr
\sin\left(K^\mu x^\mu\right)
\end{pmatrix}
,
\ee
where the amplitude $\phi$
and the numbers $K^\mu$, $\mu=1,2,3$,
serve as variational
parameters to minimize the action.

The action density, $s=S/L^2$, for an homogeneous, ferromagnetic vacuum
($K^\mu=0 ~\forall\mu$)
on a lattice $L^d$ is obtained by minimizing
\be
s_{FM}=\frac{m^2}{2}\phi^2+\frac{\lambda}{4!}\phi^4,
\ee
\be
s_{FM}^\mr{min}=
\begin{cases}
-3m^4/2\lambda&m^2<0,\cr0&m^2>0.
\end{cases}
\ee

In order to study the general case $K^\mu\neq 0$,
we shall need the eigenvector of the lattice box operator,
\be
\Delta\phi_{j,x}=-\hat P^2 \phi_{j,x},
\ee
where
\be
\hat P^2=4\sum_\mu\sin^2\left(\frac{K^\mu}{2}\right).
\label{defP}
\ee
One finds
\be
-\Delta\cK(\Delta)\phi_{j,x}=\cM\phi_{j,x},
\quad
\cM=\hat P^2(1-\hat P^2c_2+\hat P^4c_4).
\ee

Then the mean field action is
\be\label{afac}
s=\hf(m^2+\cM)\phi^2+\frac{\lambda}{4!}\phi^4,
\ee
with minimum
\be
s_{min}=-\frac{3(m^2+\cM)^2}{2\lambda}
\quad \text{at} \quad
\phi^2_{min}=-\frac{6(m^2+\cM)}{\lambda},
\label{smin}
\ee
where $m^2+\cM=m^2+\hat P^2(1-\hat P^2c_2+\hat P^4c_4)$.
This solution is valid only if $m^2+\cM\leq 0$.
The extrema of $m^2+\cM$ are reached at
\be
\hat P^2_\pm=\frac{c_2\pm\sqrt{c_2^2-3c_4}}{3c_4}
\ee
for $c_4\neq 0$, and
\be
\hat P^2_0=\frac{1}{2 c_2}
\ee
for $c_4=0$. However, $\P$ is a local minimum of Eq.~(\ref{smin}),
while $\hat P^2_-$ is a local maximum.

The local minimum $\P$ will be global or not depending
on the specific values of $c_2$ and $c_4$. Moreover, $\P$ could be
larger than 12, which is the maximum allowed value for
$\hat P^2$ from its definition Eq.~(\ref{defP}), or it could even be
an imaginary number. It is therefore necessary to carry out a
careful analysis of the absolute minima of Eq.~(\ref{smin}),
which will give us the different mean field vacua in the
$(c_2, c_4)$ plane.

\subsection{Ordered mean field phases in the $(c_2,c_4)$ plane}
\label{mainMF}

The conditions that will determine the global minimum of
Eq.~(\ref{smin}) for given $(c_2,c_4)$ are:
\begin{eqnarray}
\text{Existence of $\P$} & \Longleftrightarrow & c_4 < \frac{c_2^2}{3}
\label{P} \\
\P<12 & \Longleftrightarrow & c_4>\max\left[\frac{c_2}{36},\frac{1}{432}(-1+24
c_2)\right]
\label{P12} \\
\cM(\hat P^2=12)<0 & \Longleftrightarrow & c_4 < \frac{1}{144}(-1+12
c_2)
\label{M12} \\
\cM(\hat P^2=\P)<0 & \Longleftrightarrow & c_4 < \frac{c_2^2}{4}.
\label{MP}
\end{eqnarray}

\begin{figure}
\includegraphics[width=0.9\columnwidth]{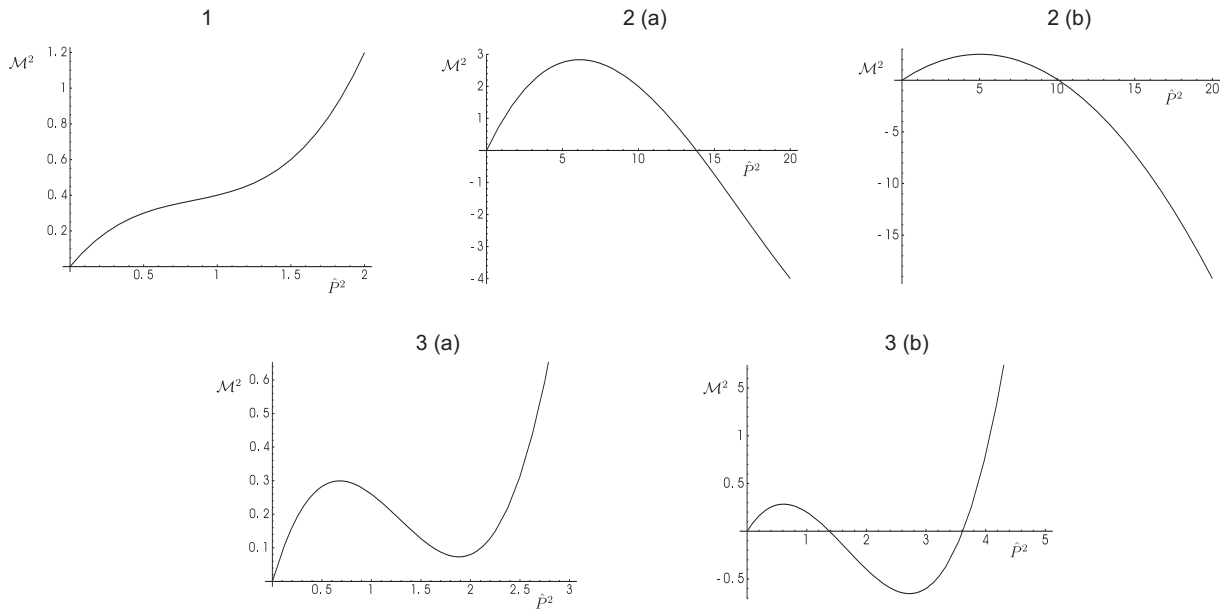}
\caption{$\cM$ as a function of $\hat{P}^2$ for different values
of $(c_2,c_4)$ corresponding to the cases signalled in the text
(from left to right, top to bottom): 1, 2\thinspace (a),
2\thinspace (b), 3\thinspace (a) and 3\thinspace (b).}
\label{fig:cases}
\end{figure}

In terms of these conditions, and supposing a value of $m^2$
such that $m^2+\cM\leq 0$, the phases (values of $K^\mu$) which minimize
Eq.~(\ref{smin}) are (see Fig.~\ref{fig:cases} for an specific
example of every case):
\begin{enumerate}
\item If condition~(\ref{P}) does \textbf{not} hold, then
  Eq.~(\ref{smin}) is minimized by
  $K^\mu=0 ~\forall \mu$:
  \textbf{ferromagnetic} vacuum.
\item If condition~(\ref{P}) \textbf{holds}, but condition~(\ref{P12})
does \textbf{not} hold, then
  \begin{enumerate}
  \item If condition~(\ref{M12}) does \textbf{not} hold, then the
    minimum is again
    $K^\mu=0 ~\forall \mu$:
    \textbf{ferromagnetic} vacuum.
  \item If condition~(\ref{M12}) \textbf{holds}, then the global minimum of
    Eq.~(\ref{smin}) is at $\P=12$ or $K^\mu=\pi ~\forall \mu$: \textbf{antiferromagnetic} vacuum.
  \end{enumerate}
\item If both condition~(\ref{P}) \textbf{and} condition~(\ref{P12})
\textbf{hold}, then
  \begin{enumerate}
  \item If condition~(\ref{MP}) does \textbf{not} hold, then
  the minimum is the \textbf{ferromagnetic} vacuum.
  \item If condition~(\ref{MP}) \textbf{holds}, then the minimum of
  Eq.~(\ref{smin}) is at the vector $K^\mu$ such that
  $\hat P^2=\P$: we will call this a \textbf{modulated} phase.
\end{enumerate}
\end{enumerate}

Note that the plot in Fig.~\ref{fig:cases}, case 3\thinspace (a), corresponds
to the dispersion relation of phonons and rotons (Landau spectrum~\cite{landau}):
we are in the ferromagnetic phase (superfluid
phase) and there are two kind of excitations, at zero and different from
zero momenta. At low energies, the curve is a straight line, corresponding to
a phonon dispersion relation, while at higher energies, the spectrum
deviates from a straight line, passing first through a maximum and then a
minimum. The excitations with energies near this minimum are called rotons.
The existence of the finite energy gap for rotons is crucial
for the superfluidity in He II. The system enters into the solid phase
when this gap tends to zero. The shape of the Landau spectrum has been confirmed
by neutron-scattering experiments carried out in several different
laboratories~\cite{explandau}.

\begin{figure}[tb]
\includegraphics[scale=0.7]{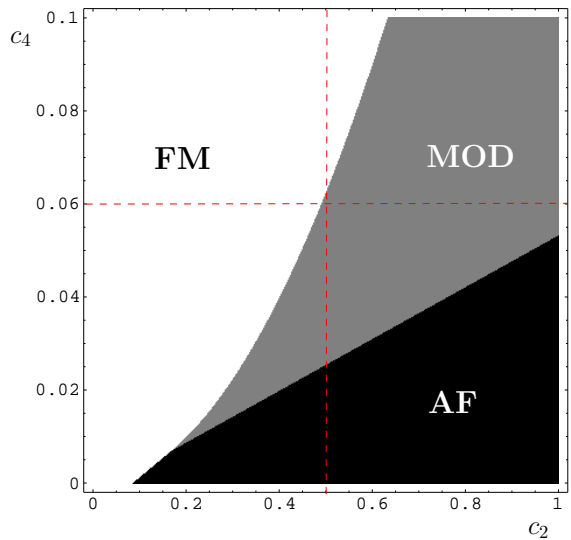}
\caption{Mean field phase diagram in the $(c_2,c_4)$ plane,
supposed that $m^2+\cM\leq 0$ at every point of the plane. The black
zone represents the AF phase, the grey zone is the MOD phase, and
the white region, the FM phase.}
\label{fig:phd}
\end{figure}

From Eqs.~\eqref{P}--\eqref{MP} and the previous discussion, we can
obtain the range of values of $(c_2,c_4)$ for which the mean field
vacuum is ferromagnetic (FM), antiferromagnetic (AF) or modulated
(MOD). The result is summarized in Table~\ref{tabla:ordvac}, and the
phase diagram which results in the $(c_2,c_4)$ plane is plotted in
Fig.~\ref{fig:phd}.

\begin{table}[tb]
\renewcommand*{\multirowsetup}{\centering}
\renewcommand{\arraystretch}{2}
\begin{tabular}{cccc} \hline\hline
$\displaystyle c_2\leq \frac{1}{12}$\smallskip & \textbf{FM} & &
$ \displaystyle\forall~c_4$ \\ \hline
\multirow{2}{1.5in}{$\displaystyle\frac{1}{12}\leq c_2\leq\frac{1}{6}$} &
\textbf{AF} & & if $\displaystyle c_4<\frac{1}{144}(-1+12 c_2)$ \\ &
\textbf{FM} & & if $\displaystyle c_4>\frac{1}{144}(-1+12 c_2)$ \smallskip \\
\hline
\multirow{3}{1.5in}{$\displaystyle\frac{1}{6}\leq c_2<\infty$} &
\textbf{AF} & & if $\displaystyle c_4<\frac{1}{432}(-1+24 c_2)$ \\ &
\textbf{MOD} & &
if $\displaystyle\frac{1}{432}(-1+24 c_2)< c_4<\frac{c_2^2}{4}$ \\ &
\textbf{FM} & & if $\displaystyle c_4>\frac{c_2^2}{4}$ \smallskip \\
\hline\hline
\end{tabular}
\caption{Mean field phases for given $(c_2,c_4)$ values, supposed
that $m^2+\cM\leq 0$.}
\label{tabla:ordvac}
\end{table}

\begin{figure}[tb]
\includegraphics[scale=0.7]{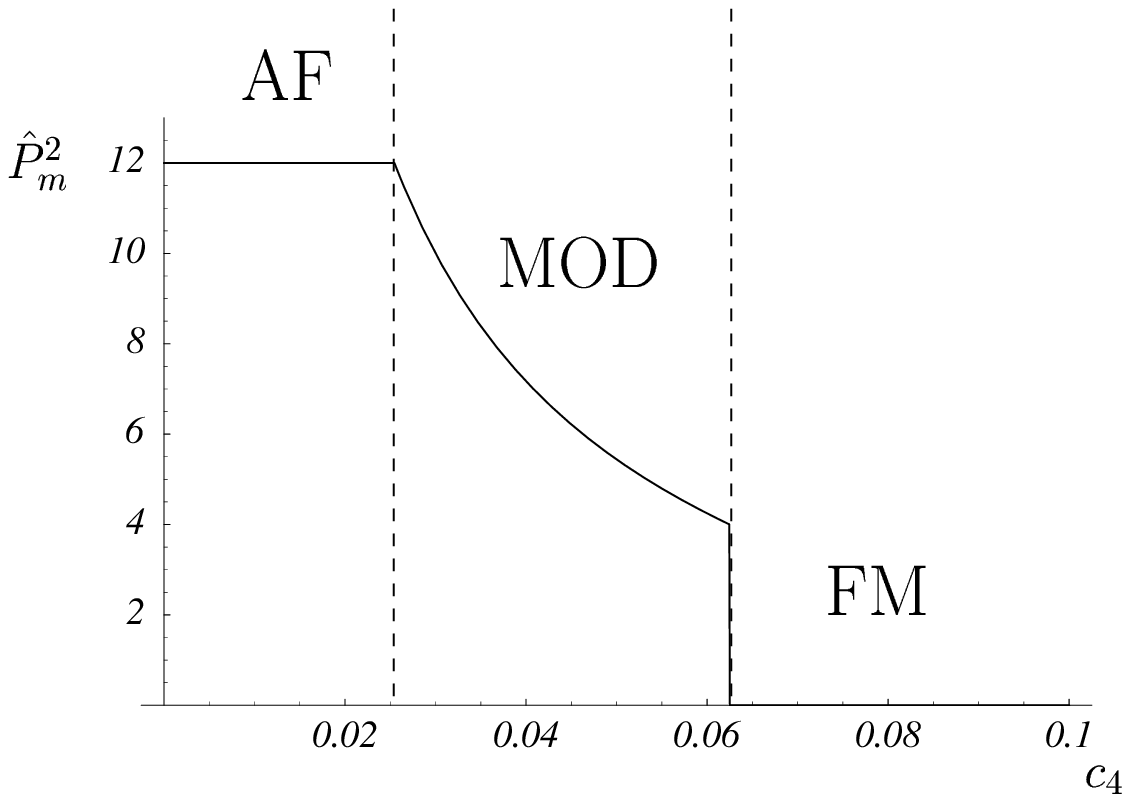}
\caption{Values of $\hat P^2$ which minimize $\cM$ at
$c_2=0.5$ as a function of $c_4$. They define the different vacua
of the mean field action, as explained in the text.}
\label{fig:vacio}
\end{figure}

Let us now consider the order of the FM-MOD and MOD-AF
transitions in this mean field approach. To do so, we will study
the variation of the minimum of the mean field action
Eq.~(\ref{smin}) along a line $c_2=\text{const}$. Let us take the
line $c_2=0.5$ which is plotted vertically in Fig.~\ref{fig:phd}.
The value of $s_{min}$ along this line depends on $\mathcal{M}^2$,
which in turn depends on the value of $\hat P^2\equiv \hat P^2_m$
which minimizes it. This value is $\hat P^2_m=12$, $\hat P_+$ and
$0$, respectively, for the antiferromagnetic, modulated and
ferromagnetic phases, and is plotted in Fig.~\ref{fig:vacio}
along the $c_2=0.5$ line. From this figure one can see that $\hat
P^2_m$ has a discontinuity when passing from the MOD to the FM
phase. This seems to suggest that $s_{min}$ is also discontinuous
and that the transition FM-MOD will be first-order. However, we
will now see that this is not the case.


The value of $s_{min}$ at the FM phase is given by (see
Eq.~(\ref{smin}), with $\mathcal{M}^2=0$):
\be
s_{min}^{FM}=-\frac{3 m^4}{2\lambda}\,.
\ee
This value is independent of $c_2$ and $c_4$. The value of
$s_{min}$ at the MOD phase is given by Eq.~(\ref{smin}), with
$\mathcal{M}^2=\mathcal{M}^2(\hat P^2_+)$. This gives
\be
s_{min}^{MOD}=-\frac{(2\, c_2^3+2(c_2^2-3c_4)^{3/2}-9\,c_2c_4-27\,c_4^2
m^2)^2}{486\, c_4^4 \lambda}\,.
\ee
However, when we evaluate this expression along the curve separating
the FM and MOD phases, $c_4=c^2_2/4$, we obtain
\be
s_{min}^{MOD}(c_2,c_4=c^2_2/4)=-\frac{3m^4}{2\lambda}=s_{min}^{FM}\,,
\ee
and there is no discontinuity in $s_{min}$ between the two phases.
Therefore, we conclude that the transition
FM-MOD is a second-order transition.

In fact we could already have arrived to this conclusion just by
a careful inspection of Fig.~\ref{fig:cases}. The transition between the
FM and MOD phases in Fig.~\ref{fig:phd} corresponds to the transition between
cases 3\thinspace (a) and 3\thinspace (b) in Fig.~\ref{fig:cases}.
In case 3\thinspace (a), $\cM(\hat P^2_+)>0$ and the minimum of the
action is at $\hat P^2_m=0$ ($\cM=0$). In case 3\thinspace (b),
$\cM(\hat P^2_+)<0$ and the minimum of the action is at $\hat P^2_+$.
At the transition, $\cM=0$ and there is no discontinuity in the action,
but there is in its derivative with respect to the parameter of variation.
The transition is second-order.

With respect to the AF-MOD transition, taking place at the curve
$c_4=(-1+24\,c_2)/432$, it is evident from Fig.~\ref{fig:vacio}
that $s_{min}$ should be continuous since $\hat P_m^2$ is continuous
at this transition. In fact, again from Eq.~(\ref{smin}),
\be
s_{min}^{AF}=-\frac{3(12-144\,c_2+1728\,c_4+m^2)^2}{2\lambda}\,,
\ee
and at the transition line
\be
s_{min}^{AF}\left(c_2,c_4=\frac{(-1+24\,c_2)}{432}\right)=
s_{min}^{MOD}\left(c_2,c_4=\frac{(-1+24\,c_2)}{432}\right)=
-\frac{3(8-48\,c_2+m^2)^2}{2\lambda}\,.
\ee
This transition corresponds to that between the cases 2\thinspace (b)
and 3\thinspace (b) in Fig.~\ref{fig:cases}. In the first case,
$\hat P^2_+>12$ and then the minimum is at $\hat P^2_m=12$,
$\cM(\hat P^2_m=12)<0$. When we approach the modulated phase then
$\hat P^2_+ \to 12$. At the transition $\hat P^2_+=12$ and
therefore $\cM$ changes in a continuous way between the two phases.
In this case the derivative of $\cM$ is also continuous.

Finally, the transition between the AF and the FM phases in the second region
of Table~\ref{tabla:ordvac} $\left(\frac{1}{12}\leq c_2\leq\frac{1}{6}\right)$,
turns out to be also continuous, in spite of the fact that $\hat P_m^2$ jumps
from $12$ to $0$ at the line $c_4=(-1+12\,c_2)/144$ since
\be
s_{min}^{AF}\left(c_2,c_4=\frac{(-1+12\,c_2)}{144}\right)=
-\frac{3 m^4}{2\lambda}=s_{min}^{FM}.
\ee
This can again be understood as the transition between cases 2\thinspace (a)
and 2\thinspace (b) in Fig.~\ref{fig:cases}, when we pass from
$\cM=0$ at the minimum of the action in the first case, to $\cM<0$ in
the second case. The change in $\cM$ is then continuous, but its derivative
is discontinuous.

In conclusion, all the transition lines in Fig.~\ref{fig:phd} turn out
to be continuous transitions. At the FM-AF and FM-MOD transitions, when
there is a jump in $\hat P^2_m$, the derivative of the action with respect
to the parameter of variation in the $(c_2,c_4)$ plane is discontinuous,
while this derivative is continuous at the AF-MOD transition.

\subsection{Complete mean field phase diagram}

In the previous analysis we have made the assumption that $m^2$
was such that $m^2+\cM\leq 0$, which guarantees that we are in
the broken phase, i.e. $\phi^2_{min}>0$ (see Eq.~\eqref{smin}).
If this condition does not hold then the minimum of the
action~\eqref{afac} is at $\phi=0$ and we are in the
paramagnetic (PM) phase. This happens at a different value of
$m^2$ depending on which ordered phase is considered:
\begin{equation}
m^2\leq -\cM \Longleftrightarrow \left\{\begin{array}{cc}
\text{FM phase} & m^2\leq 0 \\
\text{AF phase} & m^2\leq -\cM(12) \\
\text{MOD phase} & m^2 \leq -\cM(\P)\end{array}\right.
\end{equation}

\begin{figure}[tb]
\includegraphics[scale=0.7]{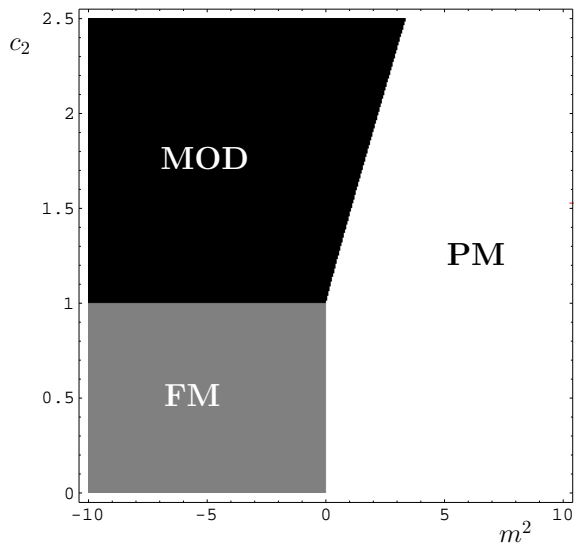}
\caption{Mean field phase diagram in the $(m^2,c_2)$ plane,
where at every point $c_4=0.25 c_2$. Here the phases
are: PM (white region), FM (grey region), and MOD (black region).}
\label{fig:hephd}
\end{figure}

In the three dimensional plane $(m^2,c_2,c_4)$ we will have then
four different phases: PM, AF, FM, and MOD. As we argued
in sections~\ref{intro} and~\ref{physint}, we are interested in
the physical region containing only FM, PM and MOD phases, which
will represent the superfluid, normal liquid and solid phases,
respectively, as a function
of two parameters: $m^2$, which would correspond to the temperature
$T$, and a combination of $c_2$ and $c_4$, which would correspond
to the pressure $P$.
Taking $c_4$ proportional to $c_2$ with an adequate slope gives us
a section in the $(c_2,c_4)$ plane represented in Fig.~\ref{fig:phd}
where the qualitatively correct phase diagram appears: a FM
phase followed by a MOD phase. The exact value of the slope
is arbitrary, and we make the choice $c_4=0.25 c_2$ so
that the transition line between the MOD and FM phases is
located at $c_2=1$.
The complete phase diagram we get in this way is shown
in Fig.~\ref{fig:hephd} in the $(m^2,c_2)$ plane.
Note that this physical region may not be stable under
renormalization group (RG) transformations. Under these transformations,
which keep the physics fixed (unmodified long distance behaviour),
the period length will appear shorter when expressed in lattice spacing
units. This means that they will in general connect the MOD phase
with an antiferromagnetic phase, but with a quite complicated
blocked action containing much more parameters than simply $c_2$ and
$c_4$. Therefore the non-physical AF phase which our simple effective
model contains will not be connected by RG transformations with the
physical region and we can safely discard this phase and its
vicinity from our analysis.

It is easy to see that the transitions from an ordered phase (FM or
MOD) to the paramagnetic phase are also continuous. The action inside the PM
phase is 0 in the mean field approximation while in the FM and the MOD phases
it is given by Eq.~\eqref{afac}.
However the transition lines to the PM phase are just the regions of the
parameter space where $m^2+\cM = 0$ and consequently the action is vanishing.


\section{Monte Carlo simulation of the model}

The preceding mean field study is qualitative only due to the
absence of fluctuations. In order to estimate this error and
to have a more reliable phase structure a numerical simulation
of the model was performed as well. We considered a three dimensional
cubic lattice of side $L$ with periodic boundary conditions and
studied the model Eq.~(\ref{laactionred}) as a function of the two
parameters $m^2$
and $c_2$, taking $\lambda=0.1$ and $c_4=0.25 c_2$ as fixed values,
in lattices $L=6,8,12,16,24$ with a Monte Carlo
simulation.
The algorithm was a standard Metropolis,
and the errors were computed with a jack-knife method.
In the largest simulations, we performed up to 50 million of
full-lattice sweeps (measuring every 5 sweeps).
We checked that the autocorrelation times were small with
respect to the number of measurements performed for every lattice size.

\subsection{Observables and the phase diagram}

\begin{figure}[tb]
\begin{center}
\includegraphics[width=0.8\columnwidth]{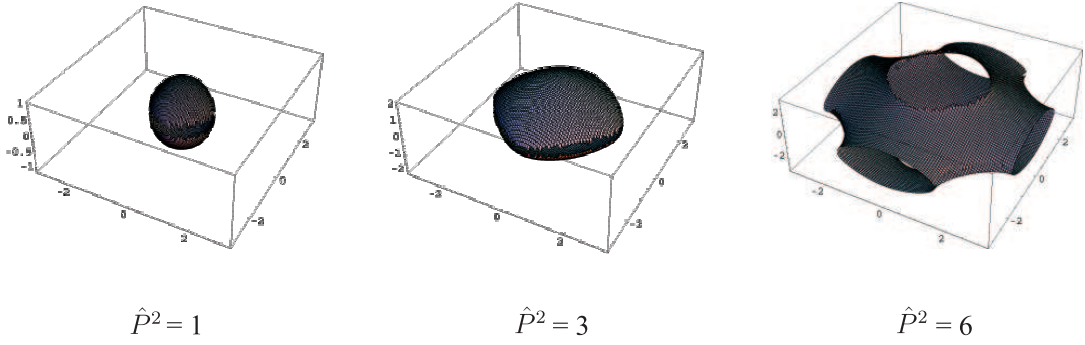}
\caption{Surfaces in $K$-space
with the same $\hat P^2$, Eq. \eqref{defP}, for three different values $\hat P^2=1,3,6$.}
\label{fig:modos}
\end{center}
\end{figure}

In order to identify the possible inhomogeneous condensates we write
the field variable $\phi_x$ in terms of its Fourier transform,
\be
\phi_x=\sum_K e^{iKx}\tilde\phi_K, \quad
\tilde \phi_K=\frac{1}{V}\sum_x e^{-iKx}\phi_x,
\end{equation}
where $Kx\equiv K^\mu x^\mu$, and there is summation over the
repeated index $\mu$.
The possible values of $K$ on each lattice direction are $(2\pi/L)n$,
with $n=0,\ldots,L-1$.
The magnetization corresponding to the wave-vector $K$
is defined as the magnitude of the Fourier coefficient,
\be
M_K=\sqrt{\lVert\tilde\phi_K\rVert^2}.
\label{Mobs}
\ee
Our mean field solutions are pure Fourier modes, i.e.~such that
$\lVert\tilde\phi_K\rVert^2=0 \ \forall K\neq\pm K_0$ for a
certain $K_0$ which minimizes the action, and we
introduce the corresponding momentum square as
\be
\hat P^2 =4 \sum_\mu \sin^2\left(\frac{K_0^\mu}{2}\right).
\ee
$\hat P^2$ is the only combination of the components of the wave-vector $K_0$
appearing in the mean field solution. But this means that all different $K$
configurations with the same value of $\hat P^2(K)$ are degenerated in energy.
The degree of this degeneration naturally may depend on the value of
$\hat P^2$. The surface in $K$-space which corresponds to a given
value of $\hat P^2$ is depicted in Fig.~\ref{fig:modos} for
three different values $\hat P^2=1,3,6$.

The configurations are not single Fourier modes of the mean field type
in the Monte Carlo simulation. A possible generalization of $\hat P^2$
is the average momentum square over all modes,
\be
\overline{\hat P^2} \equiv 4 \sum_\mu \sum_K
\lVert\tilde\phi_K\rVert^2 \sin^2\left(\frac{K^\mu}{2}\right).
\label{Pobs}
\ee

Another useful observable is the energy or in general
any of the eight terms (or any linear combination of them) appearing in
Eq.~\eqref{laactionred}. Considering the energy as a function defined on the
plane $(m^2,c_2)$, the direction of the fastest change at a transition
will be the one orthogonal to the transition line.
Since the FM-MOD and FM-PM lines are almost horizontal and vertical in
the parameter space, respectively, we considered the coefficients of $c_2$
and $m^2$ in $S$. Both linear combinations proved also to give a good signal
for the MOD-PM transition.
They can be read from Eqs.~\eq{laactionred} and~\eq{kappa1}-\eq{kappa}:
\begin{eqnarray}
S_m&\equiv& \frac{1}{2}\sum_x \phi_{j,x}\phi_{j,x},
\label{Sm} \\
S_c&\equiv& 56.25\, S_1-10.5\, S_2+0.75\, S_3-42\, S_4+9\, S_5+6\, S_6-39\, S_m\, .
\label{Sc}
\end{eqnarray}
These energies are also the appropriate ones to extrapolate the mean values of
observables in an interval around a certain simulation point by means of a
Ferrenberg-Swendsen reweighting method~\cite{FS}.
The observables Eqs.~\eqref{Pobs} and~\eqref{Mobs}, together with the
appropriate energy terms were measured in Monte Carlo simulations.

We performed two different kinds of simulations.
First we swept the whole parameter space by fixing the value of one of the
two parameters ($c_2$ or $m^2$), varying dynamically the other one
in small steps (typically $10^{-3}$) and measuring observables after a number of
iterations (around $5000$ Monte Carlo steps). This procedure allows us to
locate the transition point by means of the rapid changes experienced in the
different observables. We call this kind of
simulation an \emph{hysteresis} owing to the typical signs of metastability observed
when crossing a first-order transition (see Figs.~\ref{fig:fm-mod}
and~\ref{fig:mod-pm} below).
Once the phase diagram had been outlined we performed better statistics
simulations at fixed values of the parameters at the transition lines to
get a deeper insight into the properties of these transitions.

\begin{figure}[tb]
\includegraphics[scale=0.5]{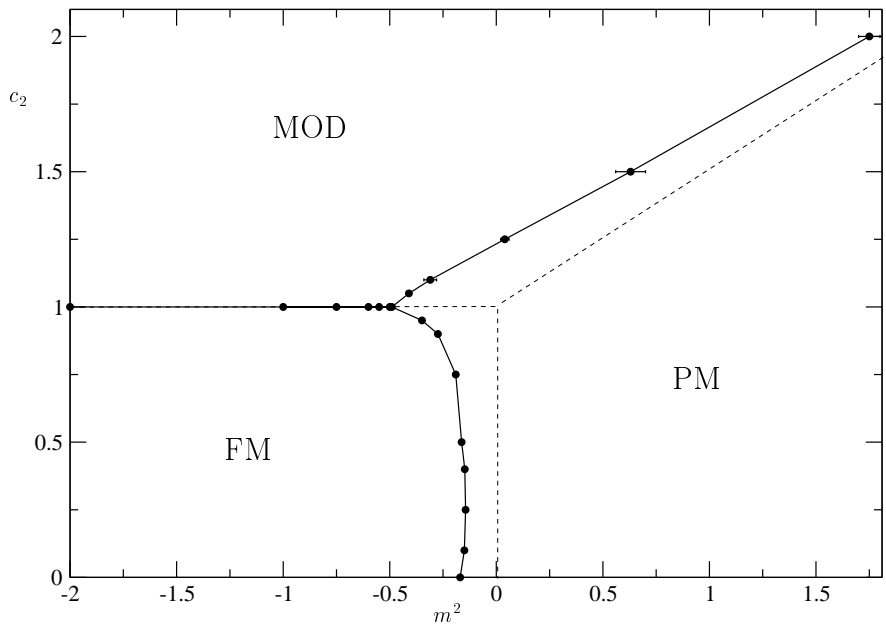}
\caption{Monte Carlo phase diagram for an $L=16$ lattice (continuous line
and dots) versus mean field phase diagram (dashed line). Error bars which
are not shown are smaller than the size of the points.}
\label{fig:MC}
\end{figure}


Figure~\ref{fig:MC} shows the observed phase
transitions in an $L=16$ simulation, together with the mean field lines.
One can see that the Monte Carlo results agree well with the mean field
phase diagram.
It is interesting to see that near the point where all phases meet,
the slopes of the different transitions are similar to those of the
helium phase diagram, see Fig.~\ref{fig:helio},
after making the
correspondence of the MOD, FM and PM phases to the solid,
liquid He II and liquid He I phases, respectively
(but note that the model does not pretend to give
the exact slopes of the transition lines).
This effective model
seems therefore to give a good description of the helium phase diagram around
the Lifshitz point when $c_2$ and $m^2$ are considered to be proportional to
the pressure and the relative temperature, respectively.
In the following section we study the
nature of the phase transitions by using a Monte Carlo simulation
and confirm that, in contrast to what we
saw in the mean field analysis, they are of the same
type as those of helium.

\subsection{Phase transitions}

\subsubsection{FM-PM transition}

The FM-PM transition is clearly second order, as we will show
by calculating its critical exponents. This behavior is expected in order
to reproduce the $\lambda$-line that separates the liquid He I and He
II states in the phase diagram of \he.

Here the order parameter is the standard magnetization $M_0$ and one can
perform the usual finite size scaling analysis in order to extract
critical exponents and critical temperatures. A good choice for the scaling
variable is the correlation length. In a finite size lattice
it can be defined using a second moment method~\cite{COOPER} as
\be
\xi=\Bigg(\frac{M^2_0/M^2_{K_{m}}-1}{4 \sin^2(\pi/L)}\Bigg)^{1/2},
\label{corr-length}
\ee
where $M^2_0$ and $M^2_{K_m}$ are defined in Eq.~\eqref{Mobs}, and
$K_m=(2\pi/L,0,0)$ is the minimum wave vector compatible with the
periodic boundary conditions.

For an operator $O$ that diverges as $|t|^{-x_O}$
where $t$ is the reduced temperature
the mean value at a temperature $T$ in a lattice of size $L$
can be written in the critical region by means of the finite-size
scaling ansatz~\cite{LIBROFSS} as
\begin{equation}
O(L,T)=L^{x_O/\nu}\left(F_O(\xi(L,T)/L)+O(L^{-\omega})\right),
\label{FSS}
\end{equation}
where $F_O$ is a smooth scaling function and $\omega$ is the
universal leading correction-to-scaling exponent.
In order to eliminate the unknown $F_O$ function we use the method of
quotients~\cite{cocientes} where one studies the behavior of the operator
of interest in two lattice sizes, $L$ and $rL$,
\begin{equation}
Q_O=O(rL,t)/O(L,t)
\end{equation}
and one chooses a value of the reduced temperature $t$ such that the
correlation-length in units of the lattice size is the same in both
lattices. This temperature can be considered as the apparent transition
point for the size $L$.
One obtains easily
\begin{equation}
\left.Q_O\right|_{Q_{\xi}=r}=r^{x_O/\nu}+O(L^{-\omega}).
\label{QUO}
\end{equation}

\begin{figure}[tb]
\includegraphics[width=0.8 \columnwidth]{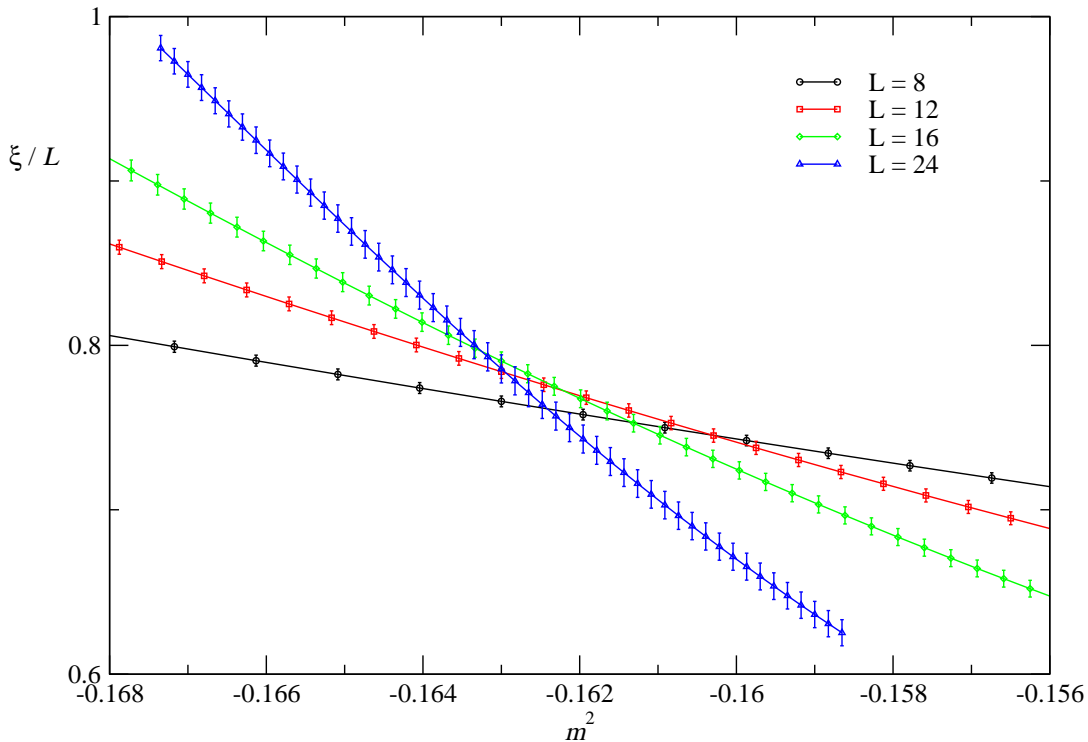}
\caption{$\xi/L$ for $c_2=0.5$ and lattice sizes $L=8,12,16,24$.}
\label{fig:cortesxi}
\end{figure}

We used the quotient method for pairs of lattices of sizes $L$
and $2L$ and determined the values of the
parameters ($m^2$,$c_2$) where the $\xi/L$ curves cut each other.
This is shown in
Fig.~\ref{fig:cortesxi} for $c_2=0.5$ and $L=8,12,16,24$. The critical
exponent $\nu$ ($\xi\sim |m^2-m^2_c|^\nu$) was measured by using
$\partial_{m^2} \xi$ ($\partial_{m^2} \xi \sim |m^2-m^2_c|^{\nu+1}$) as the observable
$O$ of the quotient method and the apparent exponents together with the
transition points are shown in Table~\ref{table-exp}.
The calculation of the scaling corrections and the exact extrapolation of the
critical point and exponents have not been carried out since our intention was
just to check that the transition belongs to the university class of the XY model in 3
dimensions (which corresponds to $c_2=0$).
More precise calculations of critical exponents would require the use
of update algorithms with smaller autocorrelation times in the vicinity of
a continuous transition than the standard Metropolis, such as a single-cluster algorithm
(for the problematics of the application of these algorithms to the standard Ginzburg-Landau
model see e.g. Ref.~\cite{bittner}). For us it is enough to confirm that
the values of the exponent
$\nu$ reported in Table~\ref{table-exp} turn out to be fully
compatible (as expected) with
that of the XY model in $3d$~\cite{XYCLASS}, $0.67155(27)$.

\begin{table}[tb]
\begin{center}
\begin{tabular}{ccccc}
\hline \hline
${\bm L}$ & &$\bm m_{c\,L}^2$ & &$\bm \nu_{L}$\\
\hline
$6$ & & $-0.1562(6)$ & & $0.663(10)$\\
$8$ & &$ -0.1613(4)$ & &$0.656(12)$\\
$12$ & &$ -0.1629(5)$ & &$0.678(17)$\\
\hline \hline
\end{tabular}
\end{center}
\caption{$m^2_{\mathrm{c}}$ determined by the intersection of the
correlation lengths measured in two lattices of sizes $L$ and $2L$ and
the apparent critical exponent $\nu$ obtained from the quotient method
applied to the same $(L,2L)$ pairs. }
\label{table-exp}
\end{table}

\subsubsection{FM-MOD transition}

\begin{figure}[tb]
\includegraphics[scale=0.7]{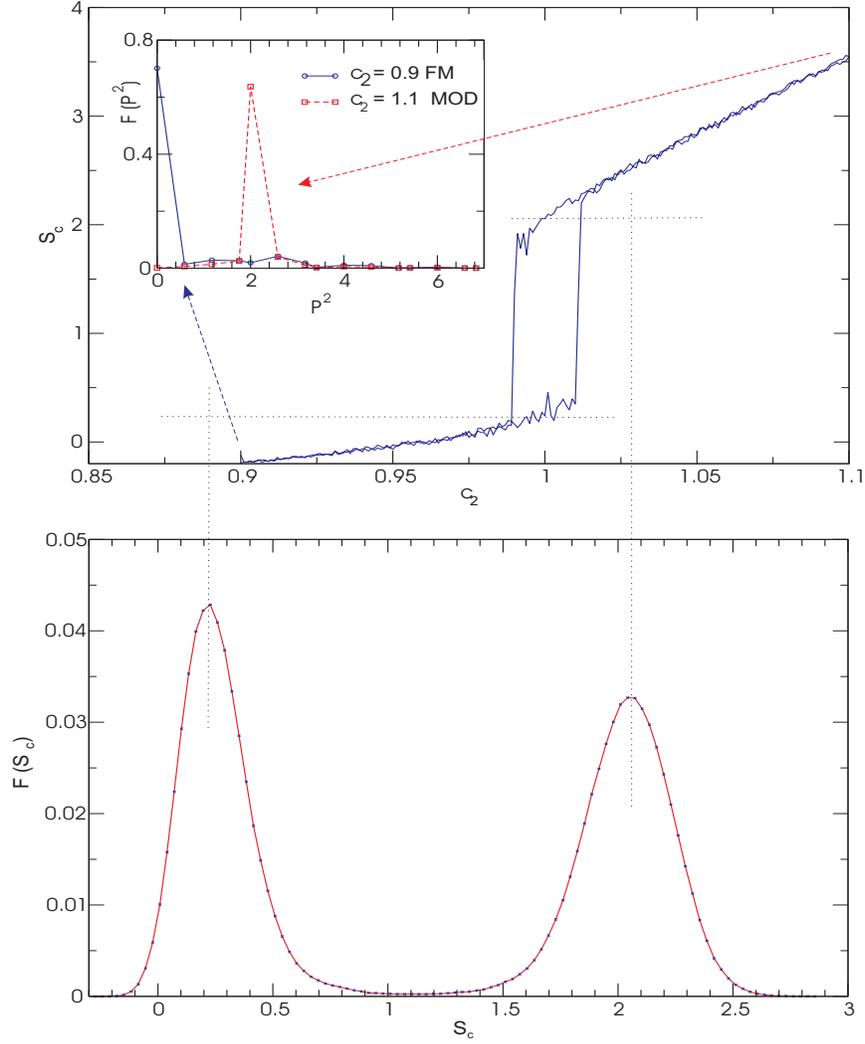}
\caption{{\bf Top:} Hysteresis cycle in $c_2$. Figure shows
$S_c$, Eq.~\eqref{Sc}, versus $c_2$ for $m^2=-0.55$ in an $L=8$ lattice. {\em
Inset:} $\overline{\p2}$ distribution, Eq.~\eqref{Pobs}, at both sides of the
transition for the same
run as in the main plot. Inside the FM phase $F(\overline{\p2})$ has a peak in
$\overline{\p2}=0$, while when crossing to the MOD phase the peak changes its
position to $\overline{\p2} \sim 2$. {\bf Bottom:} Histogram of $S_c$ for a run with
fixed parameters $c_2=1.0$, $m^2=-0.55$, in an $L=8$ lattice. Dotted lines
mark the position of the peaks of the histogram in the hysteresis
plot. The double-peak form of the histogram shows clearly the first
order character of the FM-MOD transition.}
\label{fig:fm-mod}
\end{figure}

Hysteresis-type simulations along fixed $m^2<0$ lines show metastability signs,
indicating a first-order character of the FM-MOD transition at $c_2\to 1$. This
is shown in the top part of Fig.~\ref{fig:fm-mod}.
In the small inset of this figure
the differences in the distribution of $\overline{\hat P^2}$ are shown at both
sides of the transition.
This distribution is peaked at $\overline{\hat P^2}=0$ in the
FM phase, and changes to a value clearly different from zero (around $2$) when
crossing the transition line to the MOD phase, as expected from the mean field
calculation (which gives $\hat P^2=2$ at the transition point for all values
of $m^2$).

An hysteresis might be observed for second order transitions as well
when it indicates the sudden increase of the relaxation time around
the critical point. In order to exclude this possibility we looked
into a feature characteristic of the first order transitions only,
the appearance of double-peaks in the histogram of important
observables. The histogram of the energy, shown in the bottom part
of Fig.~\ref{fig:fm-mod}, corresponds to an $L=8$ simulation at
$c_2=1.0$, $m^2=-0.55$. Such a double-peak structure is not expected
when the fluctuations around the mean field solution of
Section~\ref{mainMF} are considered at one-loop level. It is well
known that fluctuations may change the transition from second to
first-order, especially for the so-called weak first-order
transitions~\cite{weak1storder}. A first-order character is indeed
what is expected for a liquid-solid transition where there is a
finite latent heat.

Finally, Fig.~\ref{fig:MC} shows that the FM-MOD transition line obtained in the
numerical simulation is almost horizontal, here in agreement with the mean field
calculation, and also with the experimental phase diagram Fig.~\ref{fig:helio}.

\subsubsection{MOD-PM transition}

\begin{figure}[tb]
\includegraphics[scale=0.85]{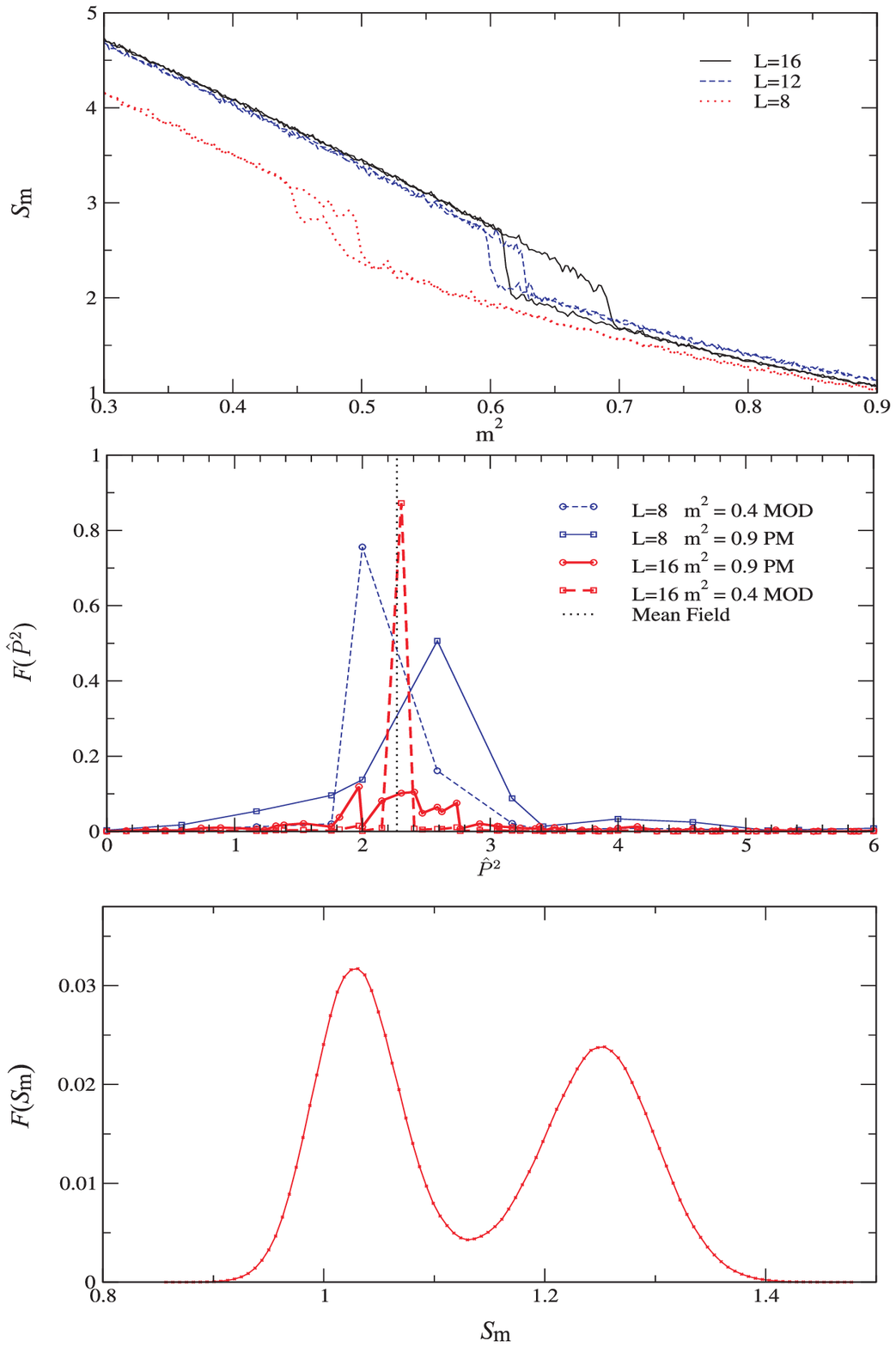}
\caption{{\bf Top:} Hysteresis cycle in $m^2$. Figure shows
$S_m$, Eq.~\eqref{Sm}, versus $m^2$ for $c_2=1.5$ for lattice sizes of
$L=8, 12, 16$. {\bf Middle:} $\overline{\p2}$ distribution at both sides of the
transition for the same run as the one in the top part
for the $L=8$ and $L=16$ lattices.
The dotted line at $\overline{\hat P^2}=2.27$ is the value of $\overline{\hat P^2}$
predicted by mean field for this value of the parameter $c_2$.
{\bf Bottom:} Histogram of $S_m$ for a run with fixed parameters
$c_2=1.5$, $m^2=0.57$ in an $L=12$ lattice. The histogram shows clearly
the first order character of the MOD-PM transition.}
\label{fig:mod-pm}
\end{figure}

The solid to ordinary liquid transition is also first-order in \he. Our
mean field solution predicted however a continuous transition. But this is
again changed by the effect of fluctuations, as the numerical simulation
reveals.

The top part of Fig.~\ref{fig:mod-pm} shows some hysteresis plots
for different lattice sizes at $c_2=1.5$. Their form shows clear
signs of a metastability at the MOD-PM transition. But notice
however the appreciable finite size dependence on the location of
the apparent transition point. This effect may be understood by
looking at the shape of the distributions of $\overline{\p2}$ of the
$L=8$ and $L=16$ lattices, shown in the middle part of
Fig.~\ref{fig:mod-pm}. The $L=16$ $\overline{\p2}$ distribution
shows a peak at the value of the mean field prediction in the
modulated phase ($2.27$ for $c_2=1.5$). The $L=8$ distribution is
however extended in a larger range with a maximum at a lower value
of $\overline{\p2}$. In fact, the spacing between Fourier modes is
$2\pi/L$, a finite number on a finite lattice, which implies that
the measured values of $\overline{\p2}$ are also discretized. The
mean field value lies between two of the allowed values for the
$L=8$ size, producing a competition between different kinds of
modulation in the system. This is of course a finite size effect
which disappears for large lattices.

The $\overline{\p2}$ distribution is spread in the paramagnetic phase.
This is clearly seen
in Fig.~\ref{fig:mod-pm}-middle for $L=16$ while the spreading is weaker for
$L=8$.

The bottom part of Fig.~\ref{fig:mod-pm} shows
the energy histogram of an $L=12$ lattice
at this transition line with the characteristic double-peak
of a first-order transition,
confirming the discontinuous character of this transition.

\section{Conclusions}

We have proposed and studied an effective field theoretical model which
is supposed to describe the \he\ phase diagram around the point where solid,
normal liquid and superfluid phases meet.
This model is a generalization of the XY model which
describes the universality class of the $\lambda$-transition and it is able to
explain the emergence of the
different phases accounting for the two-excitation dispersion relation in the
superfluid phase, and to relate the
apparition of a condensate at non-zero momentum (solid phase) with a
continuous deformation of the phonon-roton dispersion relation when
increasing the
pressure (here represented by the coefficients of the higher order derivatives)
at fixed temperature from the superfluid phase.

A mean field study, together with Monte Carlo simulations of this model,
have been performed. The numerical simulations do not modify very much the
location of the transitions in the mean field phase diagram but change in a
qualitative way the nature of the transitions obtained in the mean field
approximation. Both the form of the phase diagram and
the order of the transitions
(second-order of the XY universality class for the superfluid-normal liquid
transition, first-order for the solid-superfluid or solid-normal liquid
transitions) agree with those of \he.

This model might be used as an starting
point to study the possible apparition of
other phases, such as a ``supersolid'' phase, which, after many years of
debate~\cite{supersolid}, seems to have been experimentally
observed very recently~\cite{kim} in the \he\ system.

\begin{acknowledgments}
We thank J.L.~Alonso for useful suggestions. This work was partially
supported by Spanish MCyT FPA2001-1813 and BFM2003-08532-C02-01 research
contracts, and by DGA research group program (``Biocomputation and physics of
complex systems''). S. Jim\'enez acknowledges financial support from DGA
and from the ECHP programme, contract
HPRN-CT-2002-00307, DYGLAGEMEM.

\end{acknowledgments}

\end{document}